\documentclass[11pt,a4paper]{article}
\usepackage{jcappub}
\usepackage{bm}
\usepackage{amsmath}
\def\dd{\mathrm{d}}
\def\mcA{\mathcal{A}}
\def\mcP{\mathcal{P}}
\def\em{{\rm em}}
\def\inf{{\rm inf}}

\def\obs{{\rm obs}}

\def\GeV{{\rm GeV}}
\def\CMB{{\rm CMB}}
\def\f{{\rm f}}
\def\G{{\rm G}}
\def\reh{{\rm reh}}

\title{
Critical constraint on inflationary
magnetogenesis 
}
\author[a,b]{Tomohiro Fujita}
\author[c]{Shuichiro Yokoyama}
\affiliation[a]{Kavli Institute for the Physics and Mathematics of the
Universe (Kavli IPMU), TODIAS,  the University of Tokyo, 5-1-5
Kashiwanoha, Kashiwa, 277-8583, Japan}
\affiliation[b]{Department of Physics, University of Tokyo, Bunkyo-ku
113-0033, Japan}
\affiliation[c]{Institute for Cosmic Ray Research, University of Tokyo,
5-1-5 Kashiwa-no-Ha, Kashiwa, Chiba, 277-8582, Japan}
\emailAdd{tomohiro.fujita@ipmu.jp}
\emailAdd{shu@icrr.u-tokyo.ac.jp}

\abstract{
Recently, there are several reports that the cosmic magnetic fields on Mpc scale
in void region is larger than $\sim 10^{-15}$G with
an uncertainty of a few orders from the current blazar observations.
On the other hand, in inflationary magnetogenesis models,
additional primordial curvature perturbations are inevitably produced from
iso-curvature perturbations 
due to generated electromagnetic fields.
We explore such induced curvature perturbations 
in a model independent way
and 
obtained a severe upper bound for the energy scale of inflation
from the observed cosmic magnetic fields and 
the observed amplitude of the curvature perturbation
, as $\rho_\inf^{1/4}< 30{\rm GeV}\times(B_\obs/10^{-15}{\rm G})^{-1}$ where $B_\obs$ is the strength of the magnetic field at present.
Therefore, without a dedicated low energy inflation model or
an additional amplification of magnetic fields after inflation,
inflationary magnetogenesis on Mpc scale
is generally incompatible with CMB observations.
}

\keywords{inflation, primordial magnetic fields}
\arxivnumber{1402.0596}

\begin{document}

\begin{flushright}
ICRR-Report-669-2013-18
\\
IPMU 14-0023
\end{flushright}

\maketitle

%
%
%
\section{Introduction}

It has been known for a long time that galaxies and galactic clusters have their own magnetic fields~\cite{Wielebinski:2005,Bernet:2008qp,Bonafede:2010xg,Beck:2012}.
However, their origin is a big mystery of astronomy and 
cosmology~\cite{Giovannini:2003yn,Kandus:2010nw,Durrer:2013pga}.
Recently the generation mechanism of the magnetic fields in the universe attracts
much attention because there are several reports that magnetic fields are found even in void regions.
Such void magnetic fields could be detected by blazar observations~\cite{Neronov:1900zz, Tavecchio:2010mk, Essey:2010nd, Taylor:2011bn, Takahashi:2013uoa, Finke:2013bua} and
it is reported that their strength is larger than $\sim 10^{-15}$G with
an uncertainty of a few orders. On the other hand, the upper bound
on primordial magnetic fields could 
be also obtained 
from the cosmic microwave background (CMB) and the large scale structure (LSS) observations,
and current upper bound is roughly given by $10^{-9}$G (see, e.g. \cite{Yamazaki:2012pg,Shaw:2010ea}, and references therein)
\footnote{
Ref. \cite{Kawasaki:2012va} reported an updated constraint on a primordial magnetic field
during big bang nucleosynthesis (BBN) as $10^{-6} $G.}.
 Therefore we know there exist
the magnetic fields in the universe with the strength,
\footnote{
The upper bound is irrelevant for magnetic fields which are produced
after CMB photons are radiated.}
\begin{equation}
10^{-15}\G \lesssim B_\obs \lesssim 10^{-9}\G.
\label{Allowed range}
\end{equation}
Nevertheless, their origin is still unknown and 
no successful quantitative model is established.
If the magnetic fields are produced in the primordial universe,
they can seed the observed galactic and cluster magnetic field~\cite{Ryu:2008hi,Pakmor:2013rqa}
as well as directly explain the void magnetic fields.


As one of the mechanism of generating such cosmic magnetic fields,
``inflationary magnetogenesis" has been widely discussed. In the context of the inflationary magnetogenesis,
large scale magnetic fields, as well as the primordial curvature
perturbations, are basically generated from the quantum fluctuations.
Although many models of the generation of magnetic fields
during inflation are proposed so far~\cite{Turner:1987bw, Ratra:1991bn, Gasperini:1995dh, Davis:2000zp, Bassett:2000aw, Bamba:2003av,Enqvist:2004yy, Martin:2007ue, Ferreira:2013sqa,Garretson:1992vt,Anber:2006xt},
it is known that these inflationary magnetogenesis models suffer
from several problems, namely the strong coupling problem~\cite{Gasperini:1995dh,Demozzi:2009fu,Fujita:2012rb}, the backreaction problem~\cite{Bamba:2003av,Demozzi:2009fu,Kanno:2009ei,Fujita:2012rb}, the anisotropy problem~\cite{Bartolo:2012sd,Thorsrud:2013kya}
and the curvature perturbation problem~\cite{Suyama:2012wh,Barnaby:2012xt,Giovannini:2013rme,Ringeval:2013hfa,Fujita:2013qxa,Nurmi:2013gpa}.
In particular, the curvature perturbation problem, where
the primordial curvature perturbations
which are induced from the generated electromagnetic fields during inflation
should not exceed the observed value of CMB experiments,
gives strong 
constraints on inflationary magnetogenesis models.
For examples, in our previous paper~\cite{Fujita:2013qxa}, we have intensively studied the curvature perturbation problem by using a specific model, 
so-called the kinetic coupling model~\cite{Ratra:1991bn}, and showed that
the allowed strength of the produced magnetic fields is far weaker than
the observational lower bound given by eq.~\eqref{Allowed range}.
Ref.~\cite{Ringeval:2013hfa} have investigated the curvature perturbation problem
specifying the time evolution of the magnetic fields during inflation
as the power-law of the conformal time 
and showed limits of the amplitude of the present magnetic fields
for the monomial and the hill-top inflation models
with several reheating scenarios.

Although investigation of the constraint on inflationary magnetogenesis
in model dependent ways is important,
to discuss whether inflationary magnetogenesis is {\it really} possible or not,
model independent arguments should be also necessary.
As for such discussion,
in ref.~\cite{Suyama:2012wh} the authors have put the lower bound on the inflation energy scale $\rho_{\rm inf}$
only by requiring the production of magnetic fields with the sufficient strength $B_{\rm obs} \sim 10^{-15}$G,
but they assumed that the dominant primordial curvature perturbation
is generated during the single slow-roll inflation.
In ref.~\cite{Fujita:2012rb}, 
apart from the curvature perturbation problem, 
by requiring to escape from
the strong coupling and the backreaction problems,
the upper bound on $\rho_{\rm inf}$
has been 
put in model independent ways.

In this paper, 
we consider the curvature perturbation problem
of inflationary magnetogenesis 
in a model independent way and we do not specify the dominant contribution of the primordial curvature perturbations.
That is, our result could be also applied to the case where the dominant primordial curvature perturbation is sourced from a light scalar field other than inflaton.
We focus on the existence of the electric fields due to the time evolution of the magnetic fields
in the Friedmann-Lemaitre-Robertson-Walker (FLRW) universe and
we show that 
if one requires inflation magnetogenesis is responsible for
the generation of the observed magnetic fields and assumes no additional amplification after inflation,
the inflation energy scale is constrained 
by the curvature power spectrum $\mathcal{P}_\zeta$ 
as 
\begin{equation}
\mathcal{P}_\zeta^\obs > \mathcal{P}_\zeta^\em
\quad\Rightarrow\quad
\rho_{\rm inf}^{1/4} < 30{\rm GeV}\times 
\left(\frac{p_B}{1{\rm Mpc}^{-1}}\right)^{\frac{5}{4}} 
\left( \frac{B_{\rm obs}}{10^{-15}{\rm G}} \right)^{-1},
\label{result intro}
\end{equation}
where $\rho_\inf$ is the energy scale of inflation, $p_B > 1{\rm Mpc}^{-1}$ is the peak wave number of the void magnetic field
and $B_\obs$ is the magnetic field strength today.
Therefore, our result indicates some tension between
inflationary magnetogenesis and phenomenologies in the very early universe, e.g., genesis of the baryon or dark matter, where high energy physics are involved.

We also discuss a possible way out of our constraint.
If strong magnetic fields are produced without amplifying
electric fields, one could avoid our constraint. 
Such situation is apparently realized in a tree-level analysis of 
the so-called strong coupling regime of the kinetic coupling model
~\cite{Ratra:1991bn, Gasperini:1995dh, Bamba:2003av}.
However, since the coupling constant becomes huge in the model,
a non-perturbative analysis beyond the tree-level is required to
make the correct prediction~\cite{Demozzi:2009fu}.
Furthermore, an additional amplification or a non-adiabatic dilution
of magnetic fields after inflation can relax our constraint.
For example, if the inverse cascade works, 
the constraint is alleviated~\cite{Campanelli:2007tc, Saveliev:2013uva}.

The rest of paper is organized as follows. In section 2, we briefly review
the current lower bound on the cosmic magnetic field from the blazar observations and outline how we constrain inflationary magnetogenesis
in a model independent way.
In section 3, we derive an expression of curvature perturbations
induced by electromagnetic fields during inflation.
In section 4, the constraint on inflationary magnetogenesis is obtained.
Section 5 is devoted to a summary and discussions.
In appendix, we discuss the constraint without the assumption
of the instantaneous reheating.


\section{Basic ideas}

In this section, we briefly review the observation of the void magnetic field
and basis of our idea.
In addition, we briefly explain our approach to obtain the model independent constraint. 

\subsection{Observation of the void magnetic field}
\label{Observation of the void magnetic field}

Recently it has been reported that magnetic fields in void regions
are indirectly detected by gamma-ray observations of blazars~\cite{Neronov:1900zz, Tavecchio:2010mk, Essey:2010nd, Taylor:2011bn, Takahashi:2013uoa, Finke:2013bua}.
In such current blazar observations,
the strength and the correlation length of the magnetic field are degenerated
\footnote{In future observations, it is expected that this degeneracy will be resolved~\cite{Neronov:2013zka}.}
and hence, in the literature, the lower bound on the magnetic strength is obtained by assuming its correlation length. 
Note that if the correlation length is larger than
$\sim 1$Mpc which is roughly the mean free path of electrons and positrons in void regions, the lower bound does not depend on the correlation length.
On the other hand, in case where the correlation length is smaller than $\sim 1$Mpc,
due to the randomness of the distribution of the magnetic fields,
the effect of the magnetic fields along the line of sight
should be proportional to $(L/1 {\rm Mpc})^{1/2}$ where $L$ is a correlation length. That is, the lower bound for the strength of the magnetic fields
is proportional to $(L/1 {\rm Mpc})^{-1/2}$.
As a result, the reported lower bound for the peak strength of the magnetic field is given by~\cite{Essey:2010nd,Taylor:2011bn}
\begin{equation}
B (\eta_{\rm now} ,p_{B}) \gtrsim  10^{-15} \G
\times\left\{
\begin{array}{cc}
\left(\frac{p_B}{1 {\rm Mpc}^{-1}} \right)^{1/2}
 &  (p_B>1{\rm Mpc}^{-1})\\
 1 & (p_B<1{\rm Mpc}^{-1})
\end{array}\, \right.
,
\label{obs lower bound}
\end{equation}
where $B(\eta_{\rm now},k)$ denotes the void magnetic field at present
in Fourier space, 
$p_B$ is its peak wave number.
Note that $B(\eta_{\rm now},k)$ is assumed to has a peak at $k=p_B$
with a peak width $\Delta \ln k =\mathcal{O}(1)$ in accordance with
the definite correlation length $p_B^{-1}$.
\footnote{
A more rigorous treatment of the magnetic lower bound  is developed in the appendix of ref.~\cite{Fujita:2012rb}.}
In this paper, we focus on the case with $p_B \ge 1 {\rm Mpc}^{-1}$.

\subsection{Basis of our idea}
\label{Basics of inflationary magnetogenesis}

Let us discuss general properties of electromagnetic fields in
the FLRW universe including the inflation era.
In the FLRW universe, the Fourier transformed components of the electromagnetic fields are given in terms of the vector potential as 
\begin{equation}
E_i(\eta,\bm{k}) = -a^{-2} \partial_\eta A_i(\eta,\bm{k}),
\qquad
B_i(\eta,\bm{k}) = a^{-2} i \epsilon_{ijl}\, k_j A_l(\eta,\bm{k}),
\label{def of E and B}
\end{equation}
in the radiation gauge. Here, $a$ is the scale factor, $k$ denotes wave number, $\eta$ denotes 
conformal time and $A_i(\eta,\bm{k})$ is the vector potential in Fourier space.
Note that $B_i$ is proportional to 
$a^{-2}$ and substantially decrease as the universe expands.
For simple discussion about the strength of the electromagnetic fields,
here 
we suppress the vector legs
of $E_i, B_i$ and $A_i$.
A mathematically strict treatment including the vector legs will be shown in
the following sections.%
%

If the magnetic field is generated during inflation and it monotonically decreases by the adiabatic dilution after the inflation, the present lower bound $B(\eta_{\rm now},p_{B}) \gtrsim 10^{-15}$G can be translated into the lower bound on the
strength of the magnetic field at the end of inflation as
\begin{equation}
B(\eta_\f, p_B) \gtrsim 
10^{-15}{\rm G} \left(\frac{a_{\rm now}}{a_\f} \right)^2 
= 2\times 10^{40} {\rm G} \left(\frac{\rho_\inf^{1/4}}{10^{15}\GeV} \right)^2 ,
\label{B at inf end}
\end{equation}
where subscript f denotes the end of inflation and the instantaneous reheating is assumed for simplicity.
Therefore, to explain the observational lower bound by inflationary magnetogenesis,
strong magnetic fields should be produced during inflation.
However, the magnetic field also decreases rapidly during inflation
because of the factor $a^{-2}$. To compensate the adiabatic dilution and
produce the magnetic field effectively, 
at least the vector potential $A(\eta,p_{B})$ must be amplified faster than $a^2$ as
%
\begin{equation}
A (\eta,p_{B}) \propto |\eta|^{-n} 
\quad
(n>2).
\label{power-law}
\end{equation}
In such case where the vector potential evolves in time,
from eq.~\eqref{def of E and B} we can easily find
that 
the amplitude of the electric field should be much larger than that of the magnetic field on super-horizon scales.
From eqs.~\eqref{def of E and B} and \eqref{power-law}, we obtain
\begin{equation}
\left| \frac{E}{B} \right|= 
\left| \frac{n}{k \eta} \right| =
n e^{N_k} \gg 1
\qquad {\rm (on\ super{\text -}horizon\ scales)},
\label{hierarchy}
\end{equation}
where $N_k \equiv -\ln |k\eta|$ is 
the e-fold number measured from the end of inflation to the time at the horizon exit of the $k$ mode.
This equation means that
at the end of inflation the electric field is bigger than
the magnetic field 
whose strength is eq.~\eqref{B at inf end}
by the factor of $n e^{N_{p_{B}}}$.
Hence it is easy to imagine that including the effect of such strong electric field
into the investigation of the inflationary magnetogenesis
would give a strong constraint on the scenarios.
%
%

\subsection{Model independent approach}
\label{Model independent approach}

While most previous works specify a model of magnetogenesis 
and fix the behavior of the vector potential $A(\eta,k)$,
we assume $A (\eta,k)$ is well approximated by
a power-law of $\eta$ only for the last one e-fold of inflation.
%
%
It should be noted that 
the vector potential $A(\eta)$ can 
be a more complicated function of $\eta$ in general.
In such case, the approximation of 
the simple power-law gets worse for
considering long duration.
However, in terms of obtaining a conservative constraint in model independent approach,
it should be sufficient to focus on the contribution from the last one e-fold before the end of inflation
and assume constant $n$ during such short duration.
We also consider only the contribution from the electromagnetic fields around the peak scale $k \sim p_B$
as shown in \eqref{obs lower bound}. Of course,
in general the electromagnetic fields might have the power at the separate scales from the peak
with depending on the models
and they also give some contributions. 
Also in this respect, our constraint should be conservative, which is obtained in model independent approach. 
Thus, the key assumption of this paper for the vector potential is given by
\begin{equation}
A(\eta, k) = 
\left(\frac{\eta}{\eta_\f}\right)^{-n} 
A(\eta_\f,k)
,
\quad{\rm for}\quad
e\eta_\f \le \eta \le \eta_\f,\ k\sim p_B,\ {\rm and}\   n={\rm const}.
\label{assumption}
\end{equation}
By using this assumption for the vector potential, we will calculate
the curvature perturbation induced by the electric field
for the last one e-folding time and obtain 
the constraint by requiring that the induced curvature perturbation
is smaller than the observed value as eq.~\eqref{result intro}.

Before closing this section, it should be noted that
the constraint apparently becomes very weak 
when $A(\eta,p_B)$ significantly grows before $N=1$ 
and $A(\eta,p_B)$ is nearly constant, 
$|n| \ll 1$, for the last one e-fold.
However, in that case, we can obtain an even more stringent
constraint by considering not last one e-fold but the time when
$n\sim\mathcal{O}(1)$ before the last one e-fold. 
The details of this case will be discussed in last part of section \ref{Model Independent Constraint}.

\section{Power spectrum of Induced Curvature Perturbations}
In this section, we derive an equation that evaluates the power spectrum
of the curvature perturbation induced by the electric field during inflation.

It has been well known that the curvature perturbation on the uniform energy density hypersurface, $\zeta$,
should be constant in time on super-horizon scales when any iso-curvature component does not exist.
In case that the electromagnetic fields generated during inflation 
behave as 
the iso-curvature perturbations,
the evolution of the curvature perturbation $\zeta$ on super-horizon scales is given by
\cite{Suyama:2012wh, Nurmi:2013gpa}
\footnote{See also earlier intensive works~\cite{Giovannini:2006ph, Giovannini:2006gz, Giovannini:2006kc, Giovannini:2007aq}.}
\begin{equation}
\dot{\zeta}(t,\bm{x}) = -\frac{H(t) \delta p_{\rm nad}(t,\bm{x})}{\rho(t)+p(t)} \,,
\label{first EoM of zeta}
\end{equation}
with
the non-adiabatic pressure 
$\delta p_{\rm nad}(t,\bm{x}) \equiv \delta p(t,\bm{x}) -\frac{\dot{p}(t)}{\dot{\rho}(t)} \delta \rho(t,\bm{x})$ approximately
given by
\begin{equation}
\delta p_{\rm nad}(t,\bm{x}) \simeq {4 \over 3} \rho^{\rm em}(t,\bm{x}).
\end{equation}
Here, $H, \rho, p$ are the Hubble parameter,
total energy density and pressure, respectively,
superscript ``em" denotes that a quantity is of the electromagnetic fields and $\rho^{\rm em} = 3 p^{\rm em}$ is used.
The anisotropic stress also contributes as a source term but we conservatively ignore it~\cite{Suyama:2012wh}.
Integrating eq. \eqref{first EoM of zeta}, we obtain the 
Fourier transformed component of the curvature perturbations induced from the electromagnetic field as
\begin{align}
\zeta^\em_{\bm{k}}=
2\int dN \,\frac{\rho^\em_{\bm{k}}}{\epsilon\, \rho_\inf},
\label{first zeta}
\end{align}
where subscript ``inf" denotes that a quantity is of the inflaton, respectively.
\footnote{In the next leading order of the slow-roll parameter, the integrand in eq.~\eqref{first zeta} is multiplied by $(1-\epsilon/2)$.
This is because $p_\inf = -\rho_\inf (1-2\epsilon/3)$ and 
$\dot{p}_\inf/\dot{\rho}_\inf \simeq -1+2\epsilon/3 +\mathcal{O}(\dot{\epsilon})$ in $\delta p_{\rm nad}$.
However, since this factor does not change the order of magnitude of the integral, we ignore it.}
$N\equiv -\ln (a/a_\f)$ is the e-folding number, and
$\epsilon\equiv -\dot{H}/H^2$ is the slow-roll parameter.

The energy density of the electromagnetic field in Fourier space $\rho^\em_{\bm{k}}$
is given by
\begin{align}
\rho^\em_{\bm{k}} 
=
\frac{1}{2} \int \frac{\dd^3 p\,\dd^3 q}{(2\pi)^3}
\delta(\bm{p}+\bm{q}-\bm{k})
\left[
\bm{E}(\eta,\bm{p})\cdot \bm{E}(\eta,\bm{q})+
\bm{B}(\eta,\bm{p})\cdot \bm{B}(\eta,\bm{q})
\right]
.
\label{def of rho^em}
\end{align}
Note since $\rho^\em = (\bm{E}^2+\bm{B}^2)/2$ in the real space,
$\rho^\em_{\bm{k}}$ is written in terms of the convolution of the electromagnetic fields.
In this paper, the kinetic term of the Maxwell theory, 
$\mathcal{L} =-F^{\mu\nu} F_{\mu\nu}/4$, is assumed. 
If one consider the kinetic coupling model where 
an arbitrary function of time $I(\eta)$
is multiplied, $\mathcal{L} =-I(\eta)F^{\mu\nu} F_{\mu\nu}/4$,
eq.~\eqref{def of rho^em} is also multiplied by $I(\eta)$
(The relation between $E$ and $B$ given by eq.~\eqref{hierarchy} still holds.).
In such case, to avoid the strong coupling problem, $I(\eta)$ should be larger than unity even during inflation.
Therefore, $\rho^\em_{\bm{k}}$ is larger than eq.~\eqref{def of rho^em}
and the resultant constraint becomes tighter.
In other words, eq.~\eqref{def of rho^em} is a conservative estimate
in view of the kinetic coupling model.
Moreover, in inflationary magnetogenesis models, some interaction terms
between $A_\mu$ and other fields 
(e.g. $\mathcal{L}_{\rm int}=\frac{\phi}{M}F_{\mu\nu}\tilde{F}^{\mu\nu}$,
where $\phi$ is a pseudo-scalar field~\cite{Garretson:1992vt,Anber:2006xt})
are considered to amplify the magnetic field. In those cases,
the energy density of the interaction terms also contribute to source $\zeta$.
Nonetheless they can be conservatively ignored.
\footnote{Only if the energy density of the interaction term is negative
and it cancels the kinetic energy, our estimation becomes invalid.
But no mechanism that leads to such cancellation is found~\cite{Fujita:2012rb}.}

In FLRW universe, the power spectra of the electromagnetic fields are respectively defined as
\footnote{We consider the non-helical case where the parity is not violated.
The extension to the helical case is straightforward~\cite{Durrer:2010mq}.}
\begin{align}
\left\langle E_i(\eta,\bm{k}) E_j(\eta,\bm{k}^\prime) \right\rangle
\equiv
(2\pi)^3\delta(\bm{k}+\bm{k}^\prime)
\,\frac{1}{2}\left[\delta_{ij}-\hat{k}_i\hat{k}_j \right]
\,\frac{2\pi^2}{k^3} \mathcal{P}_E(\eta,k),
\\
\left\langle B_i(\eta,\bm{k}) B_j(\eta,\bm{k}^\prime) \right\rangle
\equiv
(2\pi)^3\delta(\bm{k}+\bm{k}^\prime)
\,\frac{1}{2}\left[\delta_{ij}-\hat{k}_i\hat{k}_j \right]
\,\frac{2\pi^2}{k^3} \mathcal{P}_B(\eta,k),
\end{align}
%
where $\langle \cdots \rangle$ denotes the vacuum expectation value.
In the radiation gauge, the vector potential $A_i(\eta,\bm{k})$ is quantized as
\begin{equation}
 A_i(\eta,\bm{k})
 = 
 \sum_{\lambda=1}^{2}  
 \epsilon_{i}^{(\lambda)}(\hat{\bm{k}}) 
 \left[ a_{\bm{k}}^{(\lambda)} \mcA_{k}(\eta) 
  + a_{-\bm{k}}^{\dag (\lambda)} \mcA_{k}^{*}(\eta) \right],
\label{mode function}
\end{equation}
where  $\epsilon_i^{(\lambda)}$ is the polarization vector,
$a^{\dagger(\lambda)}_{\bm{k}}$ and $a^{(\lambda)}_{\bm{k}}$ are respectively
creation and annihilation operators,
a hat of $\hat{\bm{k}}$ denotes the unit vector
and $\lambda$ is a polarization label.
The polarization vector $\epsilon_i^{(\lambda)}$ satisfies
$k_i \epsilon_{i}^{(\lambda)}(\hat{\bm{k}})=0,$ and
$\sum_{\lambda=1}^{2} {\epsilon}_{i}^{(\lambda)}(\hat{\bm{k}}) {\epsilon}_{j}^{(\lambda)}(-\hat{\bm{k}})
=\delta_{ij} - \hat{k}_{i} \hat{k}_{j}$
while the creation/annihilation operators satisfy a commutation relation:
$[a^{(\lambda)}_{\bm{k}},a^{\dagger(\sigma)}_{-\bm{k}'}]
= (2\pi)^3\delta(\bm{k}+\bm{k}')\delta^{\lambda \sigma}$, as usual.
From eqs.~\eqref{def of E and B} and \eqref{mode function}, the power spectra of the electromagnetic fields
can be written in terms of the mode function of the vector potential, $\mcA_k$, as
\begin{align}
\mcP_E (\eta,k) 
= \frac{k^3 |\partial_\eta \mcA_k|^2}{\pi^2 a^4},
\quad
\mcP_B (\eta,k) 
= \frac{k^5 |\mcA_k|^2}{\pi^2 a^4}.
\label{def of EM power}
\end{align}
By using those equations, one can calculate the power spectrum of the curvature perturbation
induced from the electromagnetic fields.

First, substituting eq.~\eqref{def of rho^em} into eq.~\eqref{first zeta}, we obtain
\begin{multline}
\left\langle 
\zeta^\em_{\bm{k}}
\zeta^\em_{\bm{k}'}
\right\rangle
=
\int\dd N\dd N'
\frac{1}{\epsilon\, \rho_\inf}
\frac{1}{\epsilon\, \rho_\inf}
\int\frac{\dd^3 p\,\dd^3 q \dd^3 p'\,\dd^3 q'}{(2\pi)^6}
\delta(\bm{p}+\bm{q}-\bm{k})\delta(\bm{p}'+\bm{q}'-\bm{k}')
\\\times
\left\langle
\left(
\bm{E}_{\bm{p}}\cdot\bm{E}_{\bm{q}}+
\bm{B}_{\bm{p}}\cdot\bm{B}_{\bm{q}}\right)
\left(
\bm{E}_{\bm{p}'}\cdot\bm{E}_{\bm{q}'}+
\bm{B}_{\bm{p}'}\cdot\bm{B}_{\bm{q}'}\right)
\right\rangle
.
\label{zetazeta}
\end{multline}
Here, 4-point correlation functions of the electromagnetic fields appear.
Then the 4-point correlation function of $\bm{E}$ can be computed as
\footnote{Although the complex conjugations of $\mcA_p$ are
ignored in eq.~\eqref{phase neg} for simplicity, 
they should be included like
$\left( a^{(\lambda)}_{\bm{p}}e^{i\xi_p}+a^{\dagger(\lambda)}_{\bm{-p}}e^{-i\xi_p} \right)$ where $\xi_p$ is the phase of $\mcA_p$. 
They are restored after eq.~\eqref{EE term}.
}
\begin{multline}
\left\langle
\bm{E}_{\bm{p}}\cdot\bm{E}_{\bm{q}}
\ 
\bm{E}_{\bm{p}'}\cdot\bm{E}_{\bm{q}'}
\right\rangle
=
a^{-4}(\eta) a^{-4}(\eta')
\sum_{\lambda,\sigma,\lambda',\sigma'}
\epsilon_i^{(\lambda)}(\hat{\bm{p}})
\epsilon_i^{(\sigma)}(\hat{\bm{q}})
\epsilon_j^{(\lambda')}(\hat{\bm{p}'})
\epsilon_j^{(\sigma')}(\hat{\bm{q}'})
\\\times
\partial_\eta\mcA_p(\eta)\partial_\eta\mcA_q(\eta)
\partial_{\eta'}\mcA_{p'}(\eta')\partial_{\eta'}\mcA_{q'}(\eta')
\\\times
\left\langle
\left( a^{(\lambda)}_{\bm{p}}+a^{\dagger(\lambda)}_{\bm{-p}} \right)
\left( a^{(\sigma)}_{\bm{q}}+a^{\dagger(\sigma)}_{\bm{-q}} \right)
\left( a^{(\lambda')}_{\bm{p}'}+a^{\dagger(\lambda')}_{\bm{-p}'} \right)
\left( a^{(\sigma')}_{\bm{q}'}+a^{\dagger(\sigma')}_{\bm{-q}'} \right)
\right\rangle.
\label{phase neg}
\end{multline}
Since the bracket of the annihilation/creation operators yields
$2(2\pi)^6 \delta(\bm{p}+\bm{q}') \delta(\bm{p}'+\bm{q})
\delta^{\lambda \sigma'} \delta^{\lambda' \sigma}$~\cite{Fujita:2013qxa},
performing the $q$ and $q'$ integrals by using the delta functions, one obtains
\begin{multline}
\int \frac{\dd^3 q \,\dd^3 q'}{(2\pi)^6}
\left\langle
\bm{E}_{\bm{p}}\cdot\bm{E}_{\bm{q}}
\ 
\bm{E}_{\bm{p}'}\cdot\bm{E}_{\bm{q}'}
\right\rangle
\\
=2a^{-4}(\eta) a^{-4}(\eta')
\partial_\eta\mcA_p(\eta)\partial_\eta\mcA_{p'}^*(\eta)
\partial_{\eta'}\mcA_{p'}(\eta')\partial_{\eta'}\mcA_{p}^*(\eta')
\left[
1+\left(\hat{\bm{p}}\cdot \hat{\bm{p}}'\right)^2
\right].
\label{EE term}
\end{multline}
Repeating similar calculations, one can show
\begin{align}
&\int \frac{\dd^3 q \,\dd^3 q'}{(2\pi)^6}
\left\langle
\bm{E}_{\bm{p}}\cdot\bm{E}_{\bm{q}}
\ 
\bm{B}_{\bm{p}'}\cdot\bm{B}_{\bm{q}'}
\right\rangle
\notag
\\
&\qquad\qquad\qquad
=4a^{-4}(\eta) a^{-4}(\eta')
\partial_\eta\mcA_p(\eta)\partial_\eta\mcA_{p'}^*(\eta)
\,
\mcA_{p'}(\eta')\mcA_{p}^*(\eta')
\left[\bm{p}\cdot \bm{p}'\right]^2,
\label{EB term}
\\
&\int \frac{\dd^3 q \,\dd^3 q'}{(2\pi)^6}
\left\langle
\bm{B}_{\bm{p}}\cdot\bm{B}_{\bm{q}}
\ 
\bm{B}_{\bm{p}'}\cdot\bm{B}_{\bm{q}'}
\right\rangle
\notag
\\
&\qquad\qquad
=2a^{-4}(\eta) a^{-4}(\eta')
\mcA_p(\eta)\mcA_{p'}^*(\eta)
\mcA_{p'}(\eta')\mcA_{p}^*(\eta')
\,
p^2 p'^2\left[
1+\left(\hat{\bm{p}}\cdot \hat{\bm{p}}'\right)^2
\right].
\label{BB term}
\end{align}

As we discussed in sec.~\ref{Basics of inflationary magnetogenesis}, 
the magnetic field is far smaller than the electric field 
on super-horizon. Thus we neglect the contributions that include
$\bm{B}$, namely eqs.~\eqref{EB term} and \eqref{BB term},
and  focus on eq.~\eqref{EE term}.
Note that this procedure underestimates eq.~\eqref{zetazeta}.
Substituting eq.~\eqref{EE term} into eq.~\eqref{zetazeta},
we obtain
\begin{multline}
\left\langle 
\zeta^\em_{\bm{k}}
\zeta^\em_{\bm{k}'}
\right\rangle
>
2\delta(\bm{k}+\bm{k}')
\int\dd N\dd N'
\frac{1}{\epsilon\, \rho_\inf}
\frac{1}{\epsilon\, \rho_\inf}
\int \dd^3 p\, \dd^3 p'
\delta(\bm{p}-\bm{p}'-\bm{k})
\\
\frac{\partial_\eta\mcA_p(\eta)\partial_\eta\mcA_{p'}^*(\eta)}{a^4 (\eta)}
\frac{\partial_{\eta'}\mcA_{p'}(\eta')\partial_{\eta'}\mcA_{p}^*(\eta')}{a^4 (\eta')}
\left[
1+\left(\hat{\bm{p}}\cdot \hat{\bm{p}}'\right)^2
\right].
\label{zetazeta from EE}
\end{multline}
%
By using the definition of the curvature power spectrum given as
\begin{equation}
\left\langle 
\zeta_{\bm{k}} \zeta_{\bm{k}'}
\right\rangle
=
(2\pi)^3\delta(\bm{k}+\bm{k}')
\frac{2\pi^2}{k^3}
\mathcal{P}_\zeta(k)\,,
\end{equation}
eq.~\eqref{zetazeta from EE} can be rewritten in terms of the power spectrum as
\begin{multline}
\mcP_\zeta^\em (k)
>
\frac{k^3}{2^3 \pi^5}
\int\dd N\dd N'
\frac{1}{\epsilon\, \rho_\inf}
\frac{1}{\epsilon\, \rho_\inf}
\int \dd^3 p\, \dd^3 p'
\delta(\bm{p}-\bm{p}'-\bm{k})
\\
\frac{\partial_\eta\mcA_p(\eta)\partial_\eta\mcA_{p'}^*(\eta)}{a^4 (\eta)}
\frac{\partial_{\eta'}\mcA_{p'}(\eta')\partial_{\eta'}\mcA_{p}^*(\eta')}{a^4 (\eta')}
\left[
1+\left(\hat{\bm{p}}\cdot \hat{\bm{p}}'\right)^2
\right].
\label{with 2 delta}
\end{multline}
%
This expression is a general result.

Since we consider the case
where the electromagnetic fields has a peak strength at $p_B \ge 1{\rm Mpc}^{-1}$
that are much smaller than the Planck pivot scale $k=0.05 {\rm Mpc}^{-1}$,
the delta function $\delta(\bm{p}-\bm{p}'-\bm{k})$ in the integration in terms of
$\bm{p}$ and $\bm{p}'$ can be approximated
by $\delta(\bm{p}-\bm{p}')$.
Performing the $p'$ integral with $\delta(\bm{p}-\bm{p}')$,
eq.~\eqref{with 2 delta} reads
\begin{equation}
\mcP_\zeta^\em (k)
>
\frac{k^3}{2^2 \pi^5}
\int\dd N\dd N'
\frac{1}{\epsilon\, \rho_\inf}
\frac{1}{\epsilon\, \rho_\inf}
\int^k \dd^3 p
\frac{\left|\partial_\eta\mcA_p(\eta)\right|^2}{a^4 (\eta)}
\frac{\left|\partial_{\eta'}\mcA_p(\eta')\right|^2}{a^4 (\eta')}.
\label{AA form}
\end{equation}
By using eq. \eqref{def of EM power}, 
we finally obtain
\begin{equation}
\mcP_\zeta^\em (k)
>
\frac{k^3}{4 \pi}
\int\dd N\dd N'
\frac{1}{\epsilon\, \rho_\inf(\eta)}
\frac{1}{\epsilon\, \rho_\inf(\eta')}
\int^k \frac{\dd^3 p}{p^6}\,
\mcP_E(\eta,p)\,\mcP_E(\eta',p).
\label{final form in sec 3}
\end{equation}
In the following discussion,
 we investigate the constraint on the inflationary magnetogenesis
 based on the above expression with the observed lower bound for the magnetic field given by eq. \eqref{obs lower bound}.
 
\section{Model Independent Constraint}
\label{Model Independent Constraint}

In this section, we discuss the condition that the induced curvature
power spectrum eq.~\eqref{final form in sec 3}
does not exceed the observed value.
That condition leads to a general and critical constraint on
the inflationary magnetogenesis scenarios.

\begin{figure}[tbp]
  \hspace{-2mm}
  \includegraphics[width=75mm]{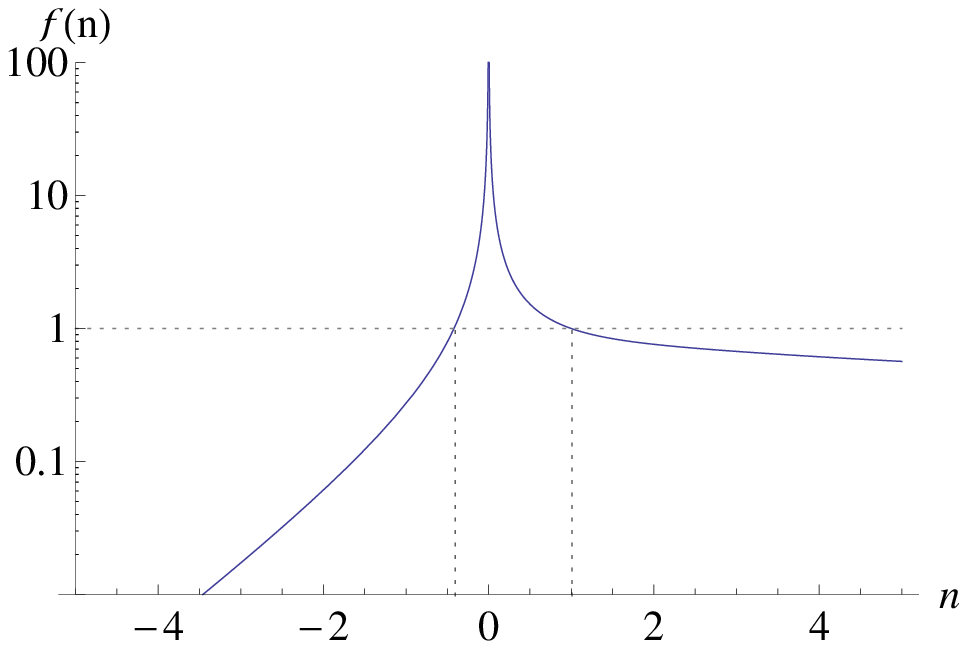}
  \hspace{5mm}
  \includegraphics[width=75mm]{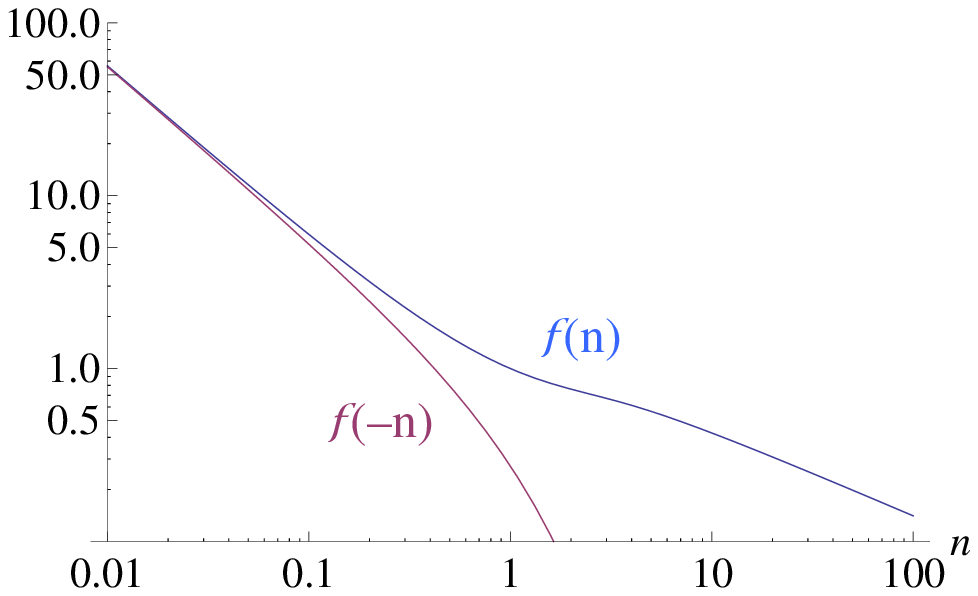}
 \caption
 {The behavior of $f(n)$ defined in eq.~\eqref{def f}.
 The left panel is the log plot while the right panel is the log-log plot.
 It is shown that $f(n)=1$ for $n=1$ and $n\approx -0.42$. One can easily see that $f(n) \le 1$ for $|n|\ge 1$ and $f(n)\gg 1$ only for $|n|\ll 1$.}
 \label{f plot}
\end{figure}

To evaluate eq.~\eqref{final form in sec 3},
we adopt the strategy outlined in sec.~\ref{Model independent approach}.
In eq.~\eqref{final form in sec 3}, the interval of the $N$ integral should be performed from the end of inflation to the time when the electric field is produced.
In the standard inflationary magnetogenesis models, the electric field is initially produced when the scale of interest exits the horizon and 
evolves until the end of inflation. Then the integration interval should be  $N=[0, \ln|k \eta_\f|^{-1}]$ where $k$ is the scale of interest and $N= \ln|k \eta_\f|^{-1}$
denotes a time at the exit of the horizon.
However, 
the time dependence of the electric field from the initial time to the end of inflation
is quite dependent on what model is considered.
Hence, as we have discussed in sec.~\ref{Model independent approach},
to obtain the conservative constraint in a model independent way,
we consider only the integration during last 1 e-folds $N=[0,1]$
and assume that the vector potential $\mcA_k(\eta)$ is a simple power-law during that period.
Moreover, we consider that 
the power spectrum of the electric field has a peak 
at a wavenumber $p_B$
which is related to the observed magnetic fields as shown in eq.~\eqref{obs lower bound}.
That is, we assume the mode function $\mcA_k(\eta)$ as
%
\begin{equation}
\mcA_k(\eta) = \left(\frac {\eta}{\eta_\f} \right)^{-n} \mcA_k(\eta_\f),
\qquad
(e\,\eta_\f \le \eta \le \eta_\f,\ k \sim p_B),
\label{Taylor}
\end{equation}
%
and by substituting eq.~\eqref{Taylor} into eq~\eqref{def of EM power}
we can relate the time dependent power spectrum of the electric field
to that of the magnetic field at the end of inflation as
\begin{equation}
\mcP_E(\eta, k) 
=
\frac{n^2}{k^2 \eta^2}
\mcP_B (\eta,k)
=
\frac{n^2}{k^2 \eta^2}
\left(\frac{\eta}{\eta_\f}\right)^{4-2n}
\mcP_B (\eta_\f,k),
\quad
(e\,\eta_\f \le \eta \le \eta_\f,\ k \sim p_B).
\label{EB relation}
\end{equation}
To connect the magnetic field at the end of inflation, $\eta_\f$,  and the present value, we assume that no amplification of the magnetic field occurs
and hence it dilutes adiabatically after inflation, $\mcP_B\propto a^{-4}$.
Although the magnetic fields on small scales vanish until today due to the plasma dissipation effect,
 such dissipation scale is about 1 AU which is much smaller than the scale of interest here and then the adiabatic dilution should be valid~\cite{Grasso:2000wj}.
For simplicity, we also assume the instantaneous reheating.%
\footnote{In appendix.~\ref{non instant reheating}
we relax this assumption for the reheating stage
and show that the similar 
constraint on the reheating energy scale $\rho_{\rm reh}$ 
can be obtained.}
Then $\mcP_B(\eta_\f,k)$ is directly connected with the present $\mcP_B(\eta_{\rm now},k)$ as
\begin{equation}
\mcP_B(\eta_\f,k)
=
\frac{\rho_\inf}{\rho_\gamma} \mcP_B (\eta_{\rm now},k) ,
\label{instant}
\end{equation}
where $\rho_\gamma \approx 5.2 \times 10^{-12} \G^2$ is the present energy density of radiation.
The lower bound for the strength of the magnetic field given by eq.~\eqref{obs lower bound} is rewritten in terms of the power spectrum as
\begin{equation}
\mcP_B(\eta_{\rm now},k)\gtrsim 
\mcP_B^{\rm obs}(p_B) \equiv 10^{-30}\G^2 
\left(
\frac{p_B}{1{\rm Mpc}^{-1}}
\right)
\qquad {\rm for}\ k\sim p_B \ge 1{\rm Mpc}^{-1}.
\label{obs lower bound 2}
\end{equation}
Substituting eqs.~\eqref{EB relation}, \eqref{instant} and \eqref{obs lower bound 2} into eq.~\eqref{final form in sec 3}, 
the $p$ integral in eq.~\eqref{final form in sec 3} reads
\begin{eqnarray}
\int \frac{\dd^3 p}{p^6}\,
\mcP_E(\eta, p)\,\mcP_E(\eta',p)
&=&
\left(\frac{\rho_\inf}{\rho_\gamma}\right)^2
\frac{n^4}{\eta^2 \eta'^2}
\left(\frac{\eta}{\eta_\f}\right)^{4-2n}
\left(\frac{\eta'}{\eta_\f}\right)^{4-2n}
\int \frac{\dd^3 p}{p^{10}}\,
\mcP_B^2(\eta_{\rm now},p)
\cr\cr
&\gtrsim&
4\pi
\left(\frac{\rho_\inf}{\rho_\gamma}\right)^2
\frac{n^4}{\eta^2 \eta'^2}
\left(\frac{\eta}{\eta_\f}\right)^{4-2n}
\left(\frac{\eta'}{\eta_\f}\right)^{4-2n}
\left(\mcP_B^\obs(p_B)\right)^2
\frac{p_B^{-7}}{7}\,,
\label{p integral}
\end{eqnarray}
where $e \eta_\f \le \eta,$ $\eta' \le\eta_\f$.
In the second line of the above equation, 
an inequality comes from the assumption that
$\mcP_B (\eta_{\rm now}, p) \simeq$ constant in $p$  for $p\sim p_B$
and $\mcP_B (\eta_{\rm now}, p) \simeq 0 $ for $p\gg p_B$ and $p\ll p_B$ while it may have a finite value
(see the discussion below eq.~\eqref{obs lower bound}). 
Then, 
as we have discussed in sec. \ref{Model independent approach},
$N$ integral within $N=[0,1]$ in eq.~\eqref{final form in sec 3} can be calculated as
\begin{equation}
\eta_\f^{2n-4} \int_0^1 \dd N\, \frac{\eta^{2-2n}}{\epsilon\, \rho_\inf}
>
\rho_\inf^{-1}\, \eta_\f^{2n-4} \int^{e\,\eta_\f}_{\eta_\f}
\dd \eta \, \eta^{1-2n}
=\rho_\inf^{-1}\, \eta_\f^{-2} \frac{1-e^{2-2n}}{2n-2},
\label{eta integral}
\end{equation}
where an inequality comes from the fact that we have used $0<\epsilon \le 1$ and $\dd N=-aH \dd\eta \simeq \frac{1}{1-\epsilon}\dd \ln \eta >\dd \ln \eta$.%
\footnote{
The factor $(1-e^{2-2n})/(2n-2)$ in eq.~\eqref{eta integral} should be replaced by 1 for $n=1$.}
We have also assumed that the energy density of the inflaton $\rho_\inf$ does not significantly vary for the last 1 e-fold.
Thus, we can obtain the conservative lower bound 
for the power spectrum of the curvature perturbations induced from the electromagnetic fields during inflation
as
\begin{equation}
\mcP_\zeta^\em (k)
>
\frac{1}{7}
\left[
n^2 \frac{1-e^{2-2n}}{2n-2}
\right]^2
\left( \frac{k}{p_B}\right)^3
e^{4N_{B}}
\left( \frac{\mcP_B^\obs}{\rho_\gamma}\right)^2
,
\label{final analytic expression}
\end{equation}
where we define $|p_B\eta_\f|^{-1}=e^{N_B}$ and $N_B$ is the e-folding number
measured between the end of inflation and a time when the $p_B$ mode exits the horizon during inflation.
$N_B$ can be written in terms of the energy density of the inflaton $\rho_{\rm inf}$
and $p_B$ as~\cite{Ferreira:2013sqa,Ferreira2}
\begin{equation}
N_B \ge
58.8- \ln\left(\frac{p_B}{H_0}\right)
+ \ln\left(\frac{\rho_\inf^{1/4}}{10^{15}\GeV}\right),
\label{Planck}
\end{equation}
where $H_0^{-1} = 4.4$Gpc is the present horizon scale and we have assumed the instantaneous reheating, and then we have
\begin{eqnarray}
\mcP_\zeta^\em (k)
>
\frac{e^{4 \times 58.8}}{7}
\left[
n^2 \frac{1-e^{2-2n}}{2n-2}
\right]^2
\left( \frac{k}{p_B}\right)^3
\left( \frac{H_0}{p_B}\right)^4
\left( \frac{\mcP_B^\obs}{\rho_\gamma}\right)^2
\left(\frac{\rho_\inf^{1/4}}{10^{15}\GeV}\right)^4
.
\end{eqnarray}
Finally, by requiring that the induced curvature perturbations
given by the above expression
 should not exceed the observed power spectrum
$\mcP_\zeta^\obs(k) =2.2\times10^{-9}$ at the Planck pivot scale $k^{-1}=20{\rm Mpc}$~\cite{Ade:2013uln}, we can obtain
the upper bound on the inflationary energy scale as
\begin{equation}
\rho_\inf^{1/4} < 
30{\rm GeV}
\left( n^2 \frac{1-e^{2-2n}}{2n-2}\right)^{-1/2}
\left(\frac{p_B}{1{\rm Mpc}^{-1}} \right)^{5/4} 
\left(\frac{B_\obs}{10^{-15}\G} \right)^{-1}.
\label{n remains}
\end{equation}
Here, we use 
$B_\obs$ given by
$\mcP_B^\obs = B_\obs^2 (p_B / 1{\rm Mpc}^{-1})$
for $p_B \ge 1{\rm Mpc}^{-1}$
which is the strength of the magnetic field measured by 
blazar observations, as shown in sec.~\ref{Observation of the void magnetic field}.
The result eq.~\eqref{n remains} depends on the parameter $n$ in the factor $f(n)$ defined by
\begin{equation}
f(n)\equiv\left( n^2 \frac{1-e^{2-2n}}{2n-2}\right)^{-1/2}.
\label{def f}
\end{equation}
$f(n)$ is plotted in fig.\ref{f plot} as a function of $n$.
In this figure, one can see $f(n)\le 1$ for $|n| \ge 1$.
Therefore $f(n)$ can be roughly replaced by 1 in  eq.~\eqref{n remains} in the case of $|n|\ge1$ and we obtain
%
%
%
%
\begin{equation}
\rho_\inf^{1/4} < 
30{\rm GeV}
\left(\frac{p_B}{1{\rm Mpc}^{-1}} \right)^{5/4} 
\left(\frac{B_\obs}{10^{-15}\G} \right)^{-1},
\qquad
(|n|\ge1).
\label{|n|>1}
\end{equation}
This is a main conclusion of this paper.

As for the case with
$|n|\ll 1$, namely $\mcA_p\simeq$ const, the constraint eq.~\eqref{n remains} seems to be relaxed because the electric field, $E\propto \partial_\eta \mcA_p$, becomes very small. Nevertheless, for $|n|\ll 1$, we can 
obtain a tighter constraint than eq.~\eqref{|n|>1} by the following argument.
This argument is based on the discussion that in order to achieve effective inflationary magnetogenesis there must exist a time when $n \sim {\cal O}(1)$
during inflation even if $|n|\ll 1$ for the last one e-fold,
as we have mentioned in the last part of section~\ref{Model independent approach}.

For the last 1 e-folding time of inflation, the magnetic power spectrum behaves as $\mcP_B \propto a^{2n-4}$ (see eqs.~\eqref{def of EM power} and \eqref{assumption}).
Thus $\mcP_B$ decreases in proportion to $a^{-4}$ for $|n|\ll 1$,
in other words, $\mcP_B$ becomes much larger as goes back in time during inflation.
On the other hand, to realize the effective production of the magnetic field during inflation,
$\mcP_B$ must significantly increase and hence $n$ should reach $\mathcal{O}(1)$ at some e-folding time $N_c$. Let us estimate the induced $\mcP_\zeta^\em$ generated
within $N=[N_c, N_c+1]$ by assuming that
$\mcA_k (\eta)$ is well approximated as
\begin{equation}
\mcA_k(\eta) = \left(\frac {\eta}{\eta_c} \right)^{-n} \mcA_k(\eta_c),
\qquad
(e\,\eta_c \le \eta \le \eta_c,\ k \sim p_B),
\end{equation}
where $\eta_c \equiv e^{N_c}\eta_\f$.
In such case, the $p$ integral in eq.~\eqref{final form in sec 3} reads
\begin{multline}
\int \frac{\dd^3 p}{p^6}\,
\mcP_E(p,N)\,\mcP_E(p,N')
\\
=
\left(\frac{\rho_\inf}{\rho_\gamma}\right)^2
\frac{n^4}{\eta^2 \eta'^2}
\left(\frac{\eta}{\eta_c}\right)^{4-2n}
\left(\frac{\eta'}{\eta_c}\right)^{4-2n}
\int \frac{\dd^3 p}{p^{10}}\,
e^{8N_c} \mcP_B^2(p,\eta_{\rm now})
\\
\gtrsim
4\pi
\left(\frac{\rho_\inf}{\rho_\gamma}\right)^2
\frac{n^4}{\eta^2 \eta'^2}
\left(\frac{\eta}{\eta_c}\right)^{4-2n}
\left(\frac{\eta'}{\eta_c}\right)^{4-2n}
\left(e^{4N_c}\mcP_B^\obs(p_B)\right)^2
\frac{p_B^{-7}}{7}\,.
\end{multline}
This equation looks similar to eq.~\eqref{p integral}.
However,
note that 
since $\mcP_B \propto a^{-4}$ for $N=[0,N_c]$, the required strength of
the magnetic field becomes large as $\mcP_B (p_B, \eta_c) = e^{4N_c} \mcP_B (p_B, \eta_\f)$ at $N_c$.
The time integration in eq.~\eqref{final form in sec 3} is given by
\begin{equation}
\eta_c^{2n-4} \int^{N_c+1}_{N_c} \dd N \eta^{2-2n}
=
\eta_c^{2n-4} \int^{e\, \eta_c}_{\eta_c} \dd \eta\, \eta^{1-2n}
=
\eta_c^{-2} \frac{1-e^{2-2n}}{2n-2}.
\end{equation}In addition, the slow-roll parameter $\epsilon$ is much smaller than unity because $N_c$ is taken to be a some time during inflation.
Thus, $\mcP_\zeta^\em (k,\eta_c)$ is bounded as
\begin{equation}
\mcP_\zeta^\em (k,\eta_c) >
\frac{1}{7}
\left[
n^2 \frac{1-e^{2-2n}}{2n-2}
\right]^2
\left( \frac{k}{p_B}\right)^3
\left( \frac{\mcP_B^\obs}{\rho_\gamma}\right)^2
e^{4N_B}\times 
\left(\frac{e^{4N_c}}{\epsilon^{2}}\right)
,
\end{equation}
where we use $e^{4N_{c}}/(p_B \eta_c)^4=e^{4N_B}$.
Note that except for the last factor, $e^{4N_c}/\epsilon^{2} \gg1$, this equation is same as eq.~\eqref{final analytic expression}. As a result, the constraint
on $\rho_\inf^{1/4}$ becomes tighter by $\sqrt{\epsilon}e^{-N_c}$ than eq.~\eqref{|n|>1} in cases where $|n| \ll 1$ for the last one e-fold of inflation, as
\begin{equation}
\rho_\inf^{1/4} < 
30{\rm GeV}
\left(\frac{p_B}{1{\rm Mpc}^{-1}} \right)^{5/4} 
\left(\frac{B_\obs}{10^{-15}\G} \right)^{-1} \sqrt{\epsilon}e^{-N_c},
\qquad
(|n|\ll 1).
\label{|n|<1}
\end{equation}
The reason why the stronger constraint is obtained can be understood as follows.
If the vector potential $\mcA_p$ stops growing and becomes constant 
during inflation ($n\sim0$),
the electric field becomes negligible.
But, at the same time, the magnetic field begins to rapidly decrease, $B\propto a^{-2}$.
To achieve the sufficient magnetic production, 
much stronger magnetic field should be generated before $\mcA_p$ stops.
Therefore the induced curvature perturbation that are  generated
right before $\mcA_p$ stops is larger than the case with $|n|\ge 1$.
\footnote{On the other hand, right before $\mcA_p$ stops,
the physical wave length of the mode $p$ is smaller
than that at the end of inflation.
Thus the hierarchy between the electric field and the magnetic field
is milder (see eq.~\eqref{hierarchy}).
Although this effect somewhat weakens the constraint,
the bound on $\rho_\inf$ becomes tighter than eq.~\eqref{|n|>1},
as a result.
}

Consequently, we conclude that eq.~\eqref{|n|>1} holds as a conservative
and general constraint on inflationary magnetogenesis for any $n$:
\begin{equation}
\rho_\inf^{1/4} < 
30{\rm GeV}
\left(\frac{p_B}{1{\rm Mpc}^{-1}} \right)^{5/4} 
\left(\frac{B_\obs}{10^{-15}\G} \right)^{-1},
\quad
(p_B \ge 1{\rm Mpc}^{-1}).
\label{final expression}
\end{equation}
%

\section{Summary and discussion}

In this paper, we show that inflationary magnetogenesis is generally
constrained as eq.~\eqref{final expression} by requiring that the curvature perturbation
induced by the electric field during inflation
should be smaller than the Planck observation
value:  $\mcP_\zeta^\obs(k) =2.2\times10^{-9}$. 
We emphasize that our argument is model independent
as we outlined in sec.~\ref{Model independent approach}.
The main result eq.~\eqref{final expression} indicates that inflationary magnetogenesis 
is under pressure in several ways. 

First, 
it is known that the reheating (thermalization) energy scale is bounded as
$\rho_{\rm reh}^{1/4} \gtrsim 10$MeV in order to achieve a successful BBN~\cite{Hannestad:2004px}.
Therefore even if eq.~\eqref{final expression} is almost saturated,
for example $\rho_{\inf}^{1/4}\sim 10$GeV,
the reheating should be quickly completed. 

Second, the generation of the observed curvature perturbation
is in danger. Eq.~\eqref{final expression}
can be translated as
\begin{equation}
H_\inf< 2\times 10^{-7} {\rm eV}
\left(\frac{p_B}{1{\rm Mpc}^{-1}} \right)^{5/2} 
\left(\frac{B_\obs}{10^{-15}\G} \right)^{-2},
\label{H constraint}
\end{equation}
where $H_\inf$ is the Hubble parameter during inflation.
For a scalar field to acquire a perturbation during inflation,
its mass should be smaller than $H_\inf$. Thus inflaton field
or a spectator field which is responsible to produce $\mcP_\zeta^\obs$
must be extremely light during inflation. During reheating era,
however, it has to quickly decay into the standard model particles to cause the BBN properly. 
Furthermore, in the case of single slow roll inflation,
eq.\eqref{H constraint} and the COBE normalization indicate
an extreme slow-roll
$\epsilon< 4\times 10^{-62}$ which demands a dedicated inflation model.
It is interesting to note that
eq.\eqref{H constraint} corresponds to the very small tensor-to-scalar ratio, $r< 7\times 10^{-61}$. Hence a detection of background gravitational waves in the future excludes
inflationary magnetogenesis.
\footnote{See ref.~\cite{Ferreira2} in which
our model-independent constraint is followed up in the light of
the BICEP2 result~\cite{Ade:2014xna}. }

Third, in such a low reheating temperature, thermal production of
the dark matter or the baryon number seems hopeless.
Since $30$GeV is accessible by colliders, 
effects beyond the standard model have been severely restricted.
To realize the dark matter production and baryogenesis,
a non-thermal mechanism like the direct decay of inflaton should be considered.

In spite of these negative implications,
since we have the observational evidence of the magnetic fields in the universe
and we are lack of a plausible magnetogenesis model,
the inflationary origin of the magnetic field is still an appealing idea.
It should be noted that we assume no amplification of the magnetic fields
after inflation to derive eq.~\eqref{final expression}.
Thus our result might imply that inflationary magnetogenesis need
an additional amplification or a non-adiabatic dilution
of magnetic fields after inflation.
If the magnetic field generated during inflation is amplified
by some mechanism like preheating process~\cite{Finelli:2000sh} or the inverse cascade~\cite{Campanelli:2007tc, Saveliev:2013uva},
the constraint is alleviated.

Another possible way out from our constraint is to 
produce a large amplitude of the vector potential 
before the horizon crossing.
It is known that, in the so-called strong coupling regime of the kinetic coupling model, the electric field is not much stronger than the magnetic
field and the backreaction and curvature perturbation
problems are evaded (if loop effects are neglected)~\cite{Demozzi:2009fu}.
This is because the vector potential $\mcA_k$ is almost
constant on super-horizon ($n\simeq 0$ in our language).
The magnetic field is produced since $\mcA_k$ has a large amplitude
at the horizon crossing due to the small kinetic function.
However, as discussed below eq.~\eqref{def of rho^em}, such model suffers
from the strong coupling problem and reliable calculations are difficult
to be done. 
If a large amplitude of a static vector potential
is realized without
the strong coupling or one can take into account the loop effects in some
non-perturbative way, sufficient magnetogenesis might be achieved.
 

\acknowledgments

We would like to thank Rajeev Kumar Jain 
and Masahiro Kawasaki for useful comments.
We are grateful to Alexander Kusenko for a useful discussion
about the observational bound on the void magnetic fields.
This work was supported by the World Premier International
Research Center Initiative (WPI Initiative), MEXT, Japan. 
T.F. and S.Y. acknowledge the support by Grant-in-Aid for JSPS Fellows
No.248160 (TF) and No. 242775 (SY).

\appendix
\section{Non instantaneous reheating}
\label{non instant reheating}

In this appendix, we relax the assumption of the instantaneous reheating.
First, it is useful to introduce the reheating parameter~\cite{Martin:2006rs,Martin:2010kz}:
\begin{equation}
R \equiv 
\left(\frac{a_\f}{a_\reh}\right)
\left(\frac{\rho_\inf}{\rho_\reh}\right)^{1/4}
=
\left(\frac{a_\reh}{a_\f}\right)^{\frac{1-3\bar{w}}{4}}
=
\left(\frac{\rho_\reh}{\rho_\inf}\right)^{\frac{1-3\bar{w}}{12(1+\bar{w})}}
,
\label{def of R}
\end{equation}
where subscript ``reh" denotes the end of reheating (thermalization) and $\bar{w}$
is the effective equation of state parameter that is
the averaged $w$ over the intermediate era between
the end of inflation and the end of thermalization.
When the assumption of the instantaneous reheating is relaxed,
two equations in 
sec.~\ref{Model Independent Constraint} are modified.
One is eq.~\eqref{instant} which should be modified as
\begin{equation}
\mcP_B(p,\eta_\f)
=
R^{-4}\, \frac{\rho_\inf}{\rho_\gamma} \mcP_B (p,\eta_{\rm now}).
\end{equation}
The other is eq.~\eqref{Planck} and it is changed as
\begin{equation}
N_B =
58.8- \ln\left(\frac{p_B}{H_0}\right)
+ \ln\left(\frac{\rho_\inf^{1/4}}{10^{15}\GeV}\right)+\ln R .
\end{equation}
Therefore the generalization to non-instantaneous reheating cases
can be taken into account by multiplying the right hand side of eq.~\eqref{final expression}
by $R$. 
If $\bar{w}>1/3$ and $R>1$,  the constraint on $\rho_\inf$ becomes milder
because the dominant component of the energy density decays faster than
the magnetic fields.

Nevertheless, it is important that $\rho_\reh^{1/4}$ can not
be bigger than the upper bound on $\rho_\inf^{1/4}$ of the instantaneous reheating case, namely eq.~\eqref{final expression}. 
Since eq.~\eqref{def of R} reads 
$\rho_\reh^{1/4}=R^{\frac{3(1+\bar{w})}{1-3\bar{w}}}\rho_\inf^{1/4}$,
$\rho_\reh^{1/4}$ can not exceed 
$R^{\frac{4}{1-3\bar{w}}}\times$ (r.h.s of eq.~\eqref{final expression}).
On the other hand, $\rho_\reh$ is smaller than $\rho_\inf$, by definition.
Except for $\bar{w}=1/3$, the constraint on $\rho_\reh^{1/4}$ can be written as
\begin{equation}
\rho_\reh^{1/4} < 
\left\{
\begin{array}{cc}
R^{\frac{4}{1-3\bar{w}}}\times30\gamma {\rm GeV}
 &  \qquad (\bar{w}>1/3,\ R^{\frac{4}{1-3\bar{w}}}<1)\\
 \rho_\inf^{1/4} < R\times 30\gamma {\rm GeV} & (\bar{w}<1/3,\ R<1)
\end{array}\, \right.,
\end{equation}
where $\gamma\equiv\left(\frac{p_B}{1{\rm Mpc}^{-1}} \right)^{5/4} 
\left(\frac{B_\obs}{10^{-15}\G} \right)^{-1}. $
Therefore the reheating (thermalization) energy scale $\rho_\reh$
is maximized in the instantaneous reheating where $R=1$ and $\rho_\inf =\rho_\reh$.


\end{document}